\documentclass[fleqn,twoside]{article}
\usepackage{espcrc2,epsfig}

\usepackage{graphicx}
\usepackage[figuresright]{rotating}

\title{Pion electromagnetic form-factor with domain wall fermions}

\author{Y. Nemoto\address[RBC]{RIKEN BNL Research Center,
        Brookhaven National Laboratory, NY 11973, USA}
	[RBC collaboration]
	\thanks{We thank RIKEN, Brookhaven National Laboratory and the U.S.\
	Department of Energy for providing the facilities essential for the
	completion of this work.}
	}
       
\begin{document}

\begin{abstract}
Motivated by recent measurements at J-Lab,
the pion electromagnetic form-factor is investigated with quenched domain 
wall fermions and a renormalization group improved
gauge action called DBW2.
We see that quark mass dependence of the form-factor with finite momentum 
transfers
is rather small.
\vspace{1pc}
\end{abstract}

\maketitle

\section{Introduction}
The pion electromagnetic (EM) or charge form-factor is one of the
simplest quantity in QCD.
Some new experiments on this quantity at relatively large
momentum transfers are under way\cite{Volmer:2000ek}.
While there exist a few lattice computations
\cite{Martinelli:1987bh},
none of them respect the chiral symmetry sufficiently well.
The new experimental data with small errors might not only impose
strong restrictions to some phenomenological models, but
give stringent test of lattice QCD.
We calculate the pion EM form-factor
in the quenched lattice QCD with domain wall fermions (DWF) with 
several momentum transfers which partially cover the above experiments.
The DWF enables us to access lighter quark masses than ever, because
it has good chiral properties which are quite important for the 
Nambu-Goldstone bosons like a pion.

\section{Simulation details}
We employ the DWF action for the quark action.
The number of sites in the fifth direction is $L_s=12$
and the domain wall height $M_5=1.8$.
Quark masses are taken to be $m_qa=$0.08, 0.06, 0.04 and 0.02
whose pion masses in the physical unit are about 760, 660, 540 and 390 MeV,
respectively.

\begin{figure}[t]
\epsfig{file=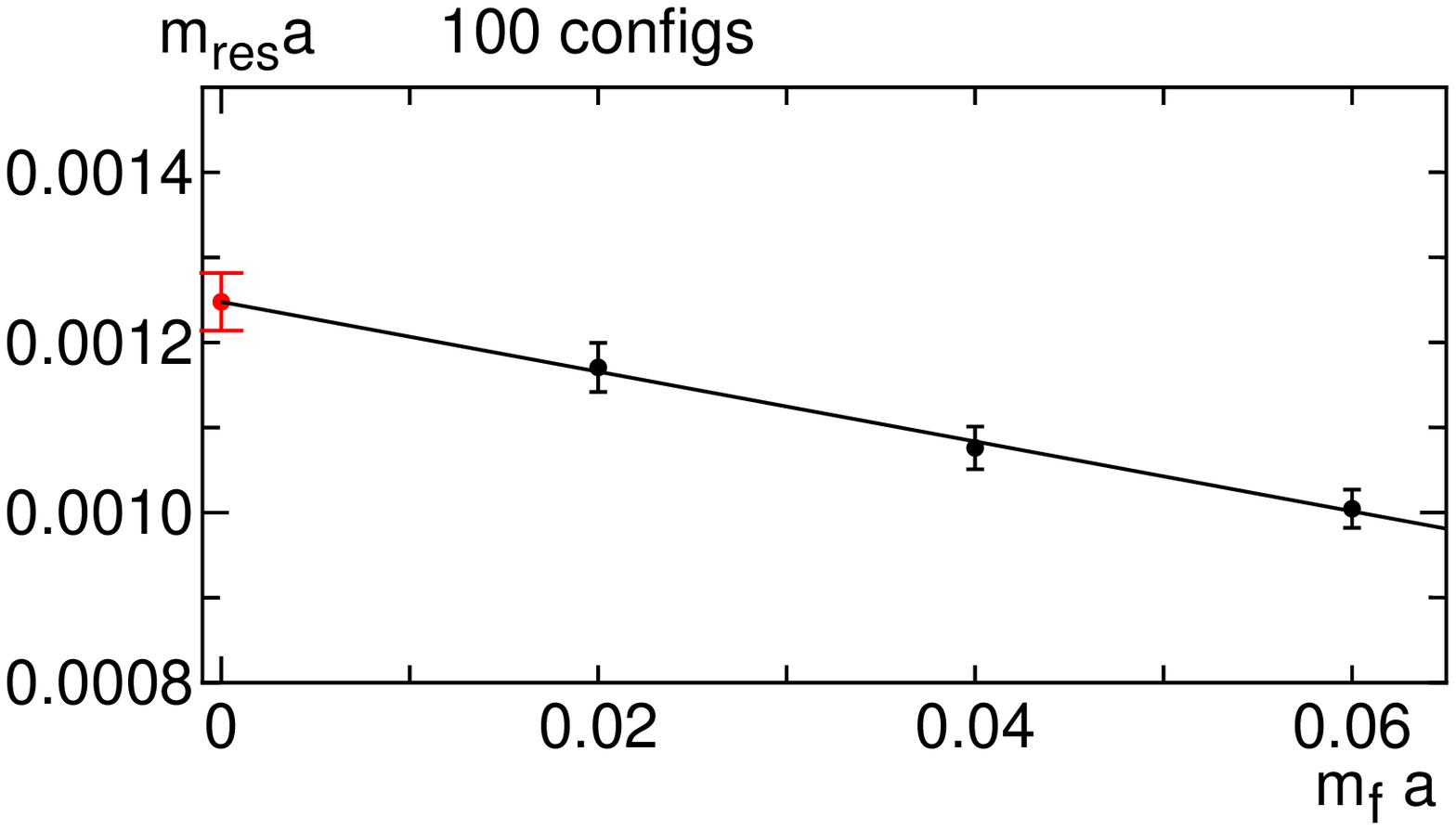, width=210pt}
\vskip -1cm
\caption{Residual masses $m_{\rm res}$ vs. quark masses $m_{\rm f}$.}
\label{fig:mres}
\vskip 3mm
\epsfig{file=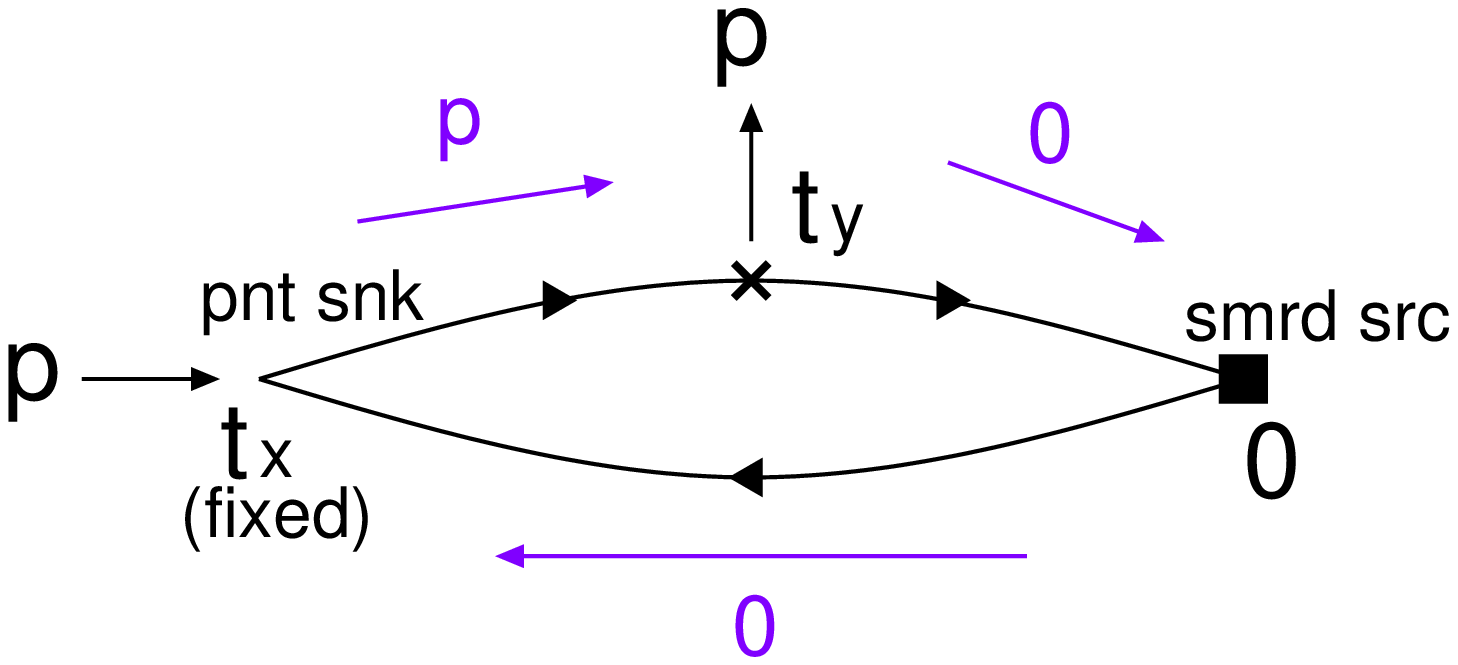, width=210pt}
\vskip -5mm
\caption{3-point function to give the pion EM form-factor.}
\label{fig:thrp}
\end{figure}

The gauge action is a renormalization group improved gauge
action called DBW2 with the lattice size $16^3\times32$,
\begin{eqnarray}
 S_G[U] &=& 
   -\frac{\beta}{3} \bigg[ (1-8 c_1) \sum_{x,\mu<\nu} P[U]_{\mu\nu}
   \nonumber \\
        & & \qquad + c_1 \sum_{x,\mu\neq\nu} R[U]_{\mu\nu} \bigg]
\end{eqnarray}
where 
$P[U]_{\mu\nu}$ is the real part of the trace of the usual $1\times1$
plaquette in the $\mu,\nu$ plane at a point $x$ and $R[U]_{\mu\nu}$
is that of the $1\times2$ rectangles in the $\mu,\nu$ plane.
$c_1=-1.4069$ is computed non-perturbatively
using the Swendsen's blocking and the Schwinger-Dyson method.
The gauge coupling is taken to be $\beta=0.87$ which gives the
lattice spacing $a^{-1} \sim 1.3$ GeV and a sufficient
spatial lattice volume $V=(2.4{\rm fm})^3$ for a pion.
The number of configuration is 100.
Some fundamental chiral properties on the DWF and DBW2 actions
are elaborated in ref.\cite{Aoki:2002vt}.
Explicit chiral symmetry breaking due to the finiteness of the
fifth direction $L_s$ is characterized by the residual mass
$m_{\rm res}$.
In fact it is shown in ref.\cite{Aoki:2002vt} that the DBW2
gauge action gives smaller $m_{\rm res}$ than those of other
(Wilson, Symanzik and Iwasaki) gauge actions.
Our present results of $m_{\rm res}$ and the linearly
extrapolated value are shown in Fig. \ref{fig:mres}.
They are much smaller than the quark masses $m_{\rm f}$.

Now we turn to the pion EM form-factor $F_{\pi}(Q^2)$ defined by
\begin{equation}
  F_{\pi}(Q^2) ( p_\mu+p'_\mu) = \langle \pi^+(p)|j_\mu|\pi^+(p')\rangle.
  \label{eq:piff}
\end{equation}
where $q^2=-Q^2=(p-p')^2$ and $j_\mu$ is the EM current.
We note $F_\pi(0)$ is the electric charge of a pion and we employ
this relation as a numerical check.
The structure of the 3-point function in eq.(\ref{eq:piff}) in our
calculation is schematically shown in Fig. \ref{fig:thrp}.
We use an appropriate smeared source and a local sink.
The separation between the source and the sink is 24 in the lattice
unit.
A momentum is injected at the sink and flows into the current insertion,
i.e., only one quark propagator has a non-zero momentum.
This method is statistically efficient for small momentum transfers.
The presently calculated momentum transfers of the
form-factor are $\vec{p}=(0,0,0),
(1,0,0)$ and $(1,1,0)$ in the unit of $2\pi/La$ with the spatial
lattice size $L$.
They correspond to physical values of $Q^2=$
0, 0.26, 0.53 GeV${}^2$, respectively.
Larger momentum transfers will be investigated in near future.
We employ the sequential source technique\cite{Bernard:1986},
when we solve the quark propagator with the finite momentum.

\begin{figure}[t]
\epsfig{file=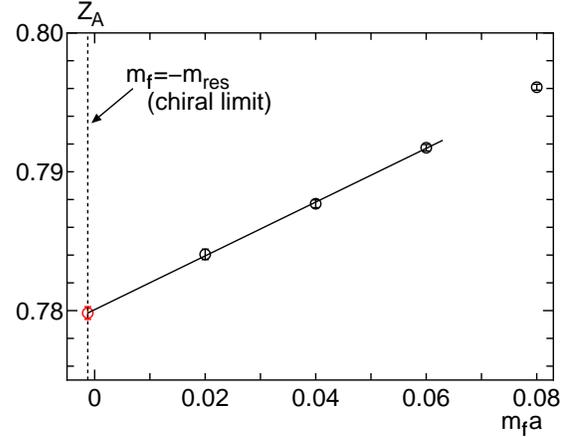, width=210pt}
\vskip -5mm
\caption{Quark mass dependence of the axial-vector renormalization 
factor $Z_A$. (100 configurations)}
\label{fig:za}
\end{figure}

The EM current taken here is the local current, 
$j_\mu=\bar{q}\gamma_\mu q$, which is not conserved on the lattice
and requires operator renormalization to compare with the matrix element
in the continuum,
\begin{equation}
  \langle \pi |j_\mu| \pi \rangle_{\rm cont} = Z_V 
  \langle \pi |j_\mu| \pi \rangle_{\rm lat}
\end{equation}
where $Z_V$ is the vector current renormalization factor.
Instead of $Z_V$, however, we use an axial-vector current
renormalization factor $Z_A$, because this quantity is so far well
investigated in the derivation of the pion decay constant
and the relation $Z_A=Z_V$ is satisfied in DWF up to $O(a^2)$
errors.
This equality is confirmed numerically in the DWF and DBW2 
action within a few percent \cite{Sasaki:2003jh}.
We show the present results of $Z_A$ in Fig. \ref{fig:za}
in the chiral limit ($m_{\rm f}=-m_{\rm res}$), 
$Z_A=0.7798(5)$.

\section{Results}
\begin{figure}[t]
\epsfig{file=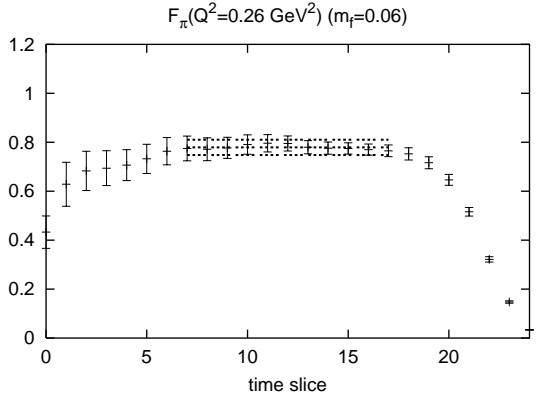, width=220pt}
\vskip -5mm
\caption{The pion EM form-factor $F_\pi(Q^2=0.26)$ 
after the renormalization with the quark mass
$m_{\rm f}a=0.06$. The vertical axis is the time slice of the
current insertion.}
\label{fig:fpi100}
\end{figure}

\begin{figure}
\epsfig{file=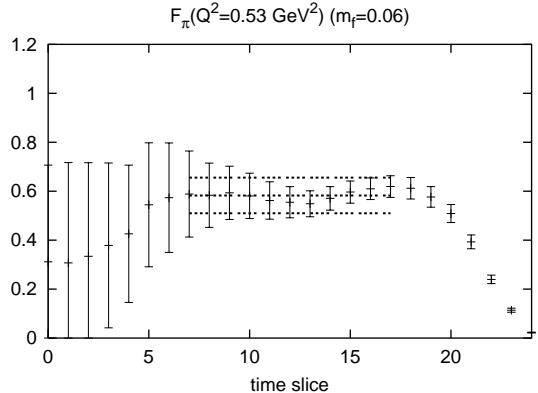, width=220pt}
\vskip -5mm
\caption{The same as Fig. \ref{fig:fpi100}, but $F_\pi(Q^2=0.53)$.}
\label{fig:fpi110}
\end{figure}

The computations were carried out by QCDSP in BNL and the Columbia
university.
We have calculated the pion EM form-factor from both the 
electric $(\langle \pi|j_{\mu=4}|\pi \rangle)$ and magnetic 
$(\langle \pi|j_{\mu=1}|\pi \rangle)$ sectors of the
3-point function and checked the consistency within the errors.
We show in Figs. \ref{fig:fpi100} and \ref{fig:fpi110} as
typical examples of calculated pion form-factors with the non-zero 
momentum transfers deduced from the matrix elements in the electric
sector ($\mu=4$). 
We see clear plateaus to give the renormalized pion form-factors
in the figures.


All the fitted results after the renormalization are summarized in Table. 
\ref{tab:res}.
We see excellent agreement with unity for $Q^2=0$, which means
that our codes and the operator renormalization $Z_A$ are quite successful.
Quark mass dependence of the form-factors with the finite momenta seems to
be rather small, although they have relatively large statistical errors.
\vskip -5mm

\begin{table}[htb]
\caption{Fitted results of the pion EM form-factor for 
each $Q^2$[GeV${}^2$] and quark mass (100 configurations).}
\begin{tabular}{c|l|l|l}
\hline
$m_{\rm f}a$ & $Q^2=0$ & $Q^2=0.26$
& $Q^2=0.53$ \\
\hline
0.08 & 1.003(19) & 0.783(24)  & 0.603(48) \\
0.06 & 1.010(22) & 0.779(31)  & 0.583(73) \\
0.04 & 1.013(27) & 0.761(44)  & 0.564(138) \\
0.02 & 1.011(34) & 0.689(77)  & n/a \\
\hline
\end{tabular}
\vskip -1cm
\label{tab:res}
\end{table}

\section{Summary and outlook}

We have calculated the pion electromagnetic
 form-factor using the quenched DWF and DBW2 
gauge action. 
Because of the good chiral and scaling properties of the action,
we have investigated lighter quark masses than ever.
The present results with small momentum transfers
show rather small quark mass dependence.

Calculations at larger momentum transfers are
necessary to obtain a functional form of the form factor to be compared
with experiments, or models such as vector meson dominance.  Such
calculations are now in progress.

\end{document}